# Immunometabolic Gatekeeping: Reconciling Peto's & the T-cell Infiltration Prognostic Paradox

Naomi Iris van den Berg[1,2,3,4,#], Matouš Elphick[1,2,5,6], Kevin Mulder[1,2], Omar Bouricha[1,2], Omid Sadeghi-Alavijeh[1,7], Xiao Fu[3,4,*], Samra Turajlic[2,1,8,#,*]

[1]The Francis Crick Institute, 1 Midland Road, London NW1 1AT, UK
[2]Cancer Research UK Manchester Institute, The University of Manchester, Wilmslow Road, Manchester, M20 4BX, UK
[3]Cancer Research UK Scotland Institute, University of Glasgow, Garscube Estate, Switchback Road, Bearsden, Glasgow G61 1BD, UK
[4]School of Cancer Sciences, University of Glasgow, Wolfson Wohl Cancer Research Centre, Bearsden, Glasgow, G61 1BD, UK
[5]Department of Bioengineering, Imperial College London, South Kensington Campus, London SW7 2AZ, UK
[6]The Institute of Cancer Research, 123 Old Brompton Road, London SW7 3RP, UK
[7]Centre for Genetics and Genomics, UCL Department of Renal Medicine, UCL Medical School, London, United Kingdom
[8]The Christie NHS Foundation Trust, Wilmslow Road, Manchester M20 4BX, UK

[#]Correspondence to:
naomi.vandenberg@crick.ac.uk, samra.turajlic@crick.ac.uk
[*]Contributed equally as senior authors.

## ABSTRACT

Classical models of cancer focus on tumour-intrinsic genetic aberrations and immune dynamics and often overlook how the metabolic environment of healthy tissues shapes tumour development and immune efficacy. Here, we propose that tissue-intrinsic metabolic intensity and waste-handling capacity act as an upstream gatekeeper of anti-tumour immunity, determining whether immune infiltration translates into effective immune function and safeguards the tissue from tumourigenesis. Across human cancers, tumours arising in high-metabolism tissues – like kidney, brain, and eye – tend to show high T cell infiltration but poor prognosis, suggesting pre-existing metabolic environments prior to malignant transformation may undermine immune function. This pattern is mirrored across species: large mammals with lower mass-specific metabolic rates (e.g., elephants, whales) accumulate fewer metabolic byproducts and show lower cancer incidence (Peto's paradox), while long-lived small mammals like bats and naked mole-rats resist tumourigenesis via suppressed glycolysis or altered hypoxia responses leading to lower metabolic rates and/or byproduct accumulation. Through integrative synthesis spanning human single-cell expression data and cross-species comparisons, we outline a framework of "immunometabolic gatekeeping," where tissues with high metabolic rate and poor waste clearance foster immune-exhausting niches even before transformation. This unifying framework reconciles multiple paradoxes in cancer biology: Peto's paradox, T cell infiltration non-prognosticity, tissue tropisms, sex-based inequalities, and size-based tipping points (e.g., the 3 cm rule in ccRCC), and suggests new principles for identifying high-risk patients and metabolic-immune combination strategies for prevention and treatment. By shifting focus from tumour-intrinsic mutations to host-tissue metabolism, this work offers a novel, integrative lens on cancer vulnerability and immune failure.

## Reframing cancer immunity through metabolic ecology

The canonical view of cancer incidence and prognosis emphasises genetic alterations and resulting cellular transformations. More recently, however, this paradigm has been challenged by an expanding body of research emphasising the role of the tumour microenvironment (TME) and host-level factors – including stromal architecture, immune contexture, and systemic metabolic health – in shaping tumour behaviour and therapeutic response[1–7]. Comparatively little attention has been given to the baseline metabolic rates of the tissues in which cancers arise, and how these may shape immune competence or dysfunction that underlie subsequent malignant transformation. Here, we suggest that the metabolic properties of specific tissues and their cellular niches from which cancers arise represent key constraints on both tumour development and immune function. For example, while immune cell dysfunction (e.g., T cell exhaustion or metabolic collapse) is widely documented in solid tumours, it remains unclear whether such dysfunction is primarily driven by local tissue conditions, tumour-derived signals, or systemic host factors such as inflammation or poor immune fitness. We introduce an immunometabolic gatekeeping framework, positing that baseline tissue metabolism governs immune competence before and after transformation. This perspective links Peto's paradox, tissue-specific differences in immune efficacy, stromal and geometric constraints, and host metabolic states into a unified model of when and where immune surveillance is likely to fail.

## The paradox of immune infiltration: when T cells do not matter

In most cancers, higher T cell infiltration correlates with improved prognosis, particularly in earlier-stage or resected tumours, where immune infiltration may reflect prior immune engagement and residual immune surveillance. For instance, in cancers such as melanoma, bladder, and colorectal – particularly following surgical resection of the primary tumour with curative intent – high T cell infiltration (e.g., 'brisk' infiltration in melanoma) is linked with reduced risk of recurrence and improved outcomes[8]. Yet exceptions to the prognosis-immune infiltration association exist: renal cell carcinoma (RCC), uveal melanoma, lower grade glioma and glioblastoma multiforme (GBM) generally show high T cell presence in the primary tumour but poor or even worse outcomes within that same cancer type[8,9]. This suggests that in some cancers, immune infiltration not only fails to confer benefit but may mark a more aggressive or immunosuppressive TME. Intriguingly, these cancers all arise in tissues with high metabolic rates and nutrient uptake per cell: the proximal tubule of the kidney, the eye, and the brain, respectively[10–13]. A cross-tissue comparison (**Table 1**) reveals an apparent inverse relationship between a tissue's metabolic intensity, gauged by single cell expression profiles, and T cell infiltration-prognosis linkage, suggesting a systemic pattern. While these associations are naturally influenced by tumour stage[14], multivariate analyses in lung adenocarcinoma (LUAD) cohorts have shown that the immune infiltrate remains an independent prognostic factor even after adjusting for stage, age, and sex[8], indicating that immune contexture contributes unique prognostic information beyond conventional clinical variables.

High metabolic rates in solid tissues can create a local imbalance between the production and diffusion rates of metabolic byproducts, resulting in their accumulation, such as reactive oxygen species (ROS) generated via oxidative phosphorylation (OXPHOS) or lactate and protons generated via glycolysis. Separately, in metabolically intense tissues where glucose and/or fatty acid catabolism compete for oxygen, oxygen scarcity often triggers a metabolic shift toward glycolysis. This adaptation supports rapid ATP production and biosynthetic capacity despite limited oxygen availability, and is a well-documented feature of developing embryonic tissues, stem cells, activated/proliferating immune cells (i.e., T cells, M1-like macrophages[15], dendritic cells[16]) and highly active tissues such as the brain, eye and kidney[17–19]).

**Table 1. Comparison of T cell infiltration, prognostic impact, and baseline tissue metabolic activity across 15 normal tissues associated with cancer initiation.**

Comparison of T cell infiltration, prognostic impact, and baseline metabolic features across 15 normal tissues associated with cancer initiation. Tissues are ordered by the strength of the $CD8^+$ T cell infiltration-prognosis relationship, based on the cross-cohort meta-analysis of resected tumours by Bruni et al. (2020)[8], consistent with other pan-cancer studies[8,]. Higher intrinsic metabolic intensity – estimated from stromal GLUT1* (SLC2A1) expression, mitochondrial respiration (complex IV), and a panel of pH-homeostatic genes – tends to associate with a weaker or even adverse prognostic impact of T cell infiltration. TOX expression in normal-tissue T cells was used as a representative exhaustion marker because TOX encodes the master transcriptional regulator that stabilises the exhausted T cell fate, providing a robust, fate-level index of chronic exhaustion pressure across tissues. Single cell expression values were derived from normal human fibroblasts, endothelial cells, and T cells in CELLxGENE[20] and processed as detailed in **Supporting Materials IA**. Stromal to T cell ratios reflect relative cell representation in the underlying scRNA-seq datasets after QC filtering. Mitochondrial respiration values were derived from healthy young mice (**Supporting Materials IB**). 'Poor' = predominantly adverse or null prognostic effect; 'Poor, very'/'Good, very' = uniformly adverse/beneficial.

| **Normal tissue** (cancer type) | CD8+ Infiltr. → Progn. [8] | T reg** Infiltr. → Progn. [8] | SLC2A1 (GLUT1) expr. fibroblasts & endothelial cells in normal human tissue[20] | pH homeostasis expr. fibroblasts & endothelial cells in normal human tissue[20] | TOX expr. T cells in normal human tissue | Mitochondrial resp. average (complex IV) in mice [12] | **No of publ.**/ Ratio of endo + fib. cells vs. T cells[20] | Notes |
|---|---|---|---|---|---|---|---|---|
| **Kidney** (Renal cell carcinoma) | Poor, very | Poor, very | High | High | High | High, very | **13** / 3:1 | The renal proximal tubule, where many RCCs originate, is among the highest in mitochondrial density and oxygen use[21] |
| **Brain** (glioma) | Poor | Poor, very | High, very | High | High, very | High, very | **28** / 56:1 | |
| **Eye** (uveal melanoma) | Poor[22] | Poor | High | Intermediate | High | Int-High | **12** / 15:1 | Not included in [8] |
| **Oesophageal** | Poor | Poor | High, very | High, very | Int-High | NA | **5** / 9:1 | |
| **Prostate gland** (prostate) | Int-Poor | Poor, very | Int-High | High | Intermediate | NA | **5** / 2:1 | |
| **Stomach** (gastric) | Intermediate | Int-Poor | Int-High | Intermediate | Intermediate | High | **6**/ 1:1 | |
| **Pancreas** (pancreatic ductal adenocarcinoma) | Int-Good | Poor, very | High | High, very | Int-High | Intermediate | **10** / 8:1 | Ductal cells have high metabolic rates[23]. Calorie (and thus metabolic rate) restrictions reduce PDAC tumour growth *in vivo*[23] |
| **Lung** (lung cancers, incl. adenocarcinoma) | Int-Good | Poor, very | Int-High | High | High | Int-High | **26** / 2:1 | |
| **Head and neck** | Good | Int-Poor | NA | NA | NA | NA | | |
| **Breast** | Good, very | Poor | Intermediate | Intermediate | Int-High | NA | **6** / 9:1 | |
| **Liver** (hepatocellular carcinoma) | Good, very | Poor | Int-Low | Intermediate | Low | High | **13** / 1:1 | |
| **Fallopian tubes** (ovarian[24]) | Good, very | Intermediate | Int-Low | Low | Low | Int-Low | **5** / 3:1 | |
| **Skin** (melanoma) | Good, very | Intermediate | Int-Low | Low | Low | Low | **10** / 3:1 | |
| **Colon** (colorectal) | Good, very | Good | Low | Int-Low | Int-Low | Low | **13** / 1:1 | |
| **Bladder** (bladder organ, urinary bladder) | Good, very | Good, very | Low | Low | Int-Low | NA | **3** / 3:1 | 2 scRNAseq datasets for 'bladder organ', 1 for urinary bladder |

*Experimental models support the view that GLUT1 acts as a rate-limiting gatekeeper of malignant transformation: in HER2-driven mammary models, GLUT1 deletion prevents tumour initiation despite not affecting growth once tumours are established, underscoring glucose uptake as essential for transformation rather than metastatic progression[25].

**Note that associations in **Table 1** reflect aggregate rather than subtype-specific effects of T reg infiltration.

While glycolysis contributes to acidosis via lactate and proton buildup, oxidative metabolism can also acidify the microenvironment through mitochondrial $CO_2$ production[26]. As $CO_2$ diffuses into the cytosol, it combines with water to form carbonic acid ($H_2CO_3$), which dissociates into bicarbonate and protons. The protons can then be actively exported into the extracellular space via proton transporters or ion exchangers. In tissues with limited perfusion or structural barriers to diffusion, where buffering systems can get (transiently) overwhelmed, this proton efflux can lead to localised extracellular acidification, independent of lactate. A recent study in diverse human and mouse cell types (including cancer cell lines, primary T cells, and isolated mitochondria) found that as extracellular lactate accumulates, it can enter the mitochondrial matrix and stimulate electron transport chain (ETC) activity independently of its metabolism[27]. This not only highlights crosstalk between metabolic pathways but also raises the possibility that lactate may promote further reactive oxygen species (ROS) production, reinforcing local metabolic stress. Together, the buildup of metabolic byproducts (e.g., ROS, lactate, carbonic acid), protons, protein aggregates and stress granules may reinforce a metabolically hostile niche that disables infiltrating immune cells[28–31], contributing to immune dysfunction even in the presence of high infiltration.

We thus propose that tissues with high metabolic activity are predisposed to produce byproduct-rich, nutrient-competing, immunosuppressive environments upon or before transformation, thereby decoupling immune infiltration from effective immune function; either by exhausting effector T cells or by skewing infiltrates toward anti-inflammatory phenotypes. **Fig. 1** outlines this immunometabolic gatekeeping framework and how it expands the classical link between metabolism and tumourigenesis.

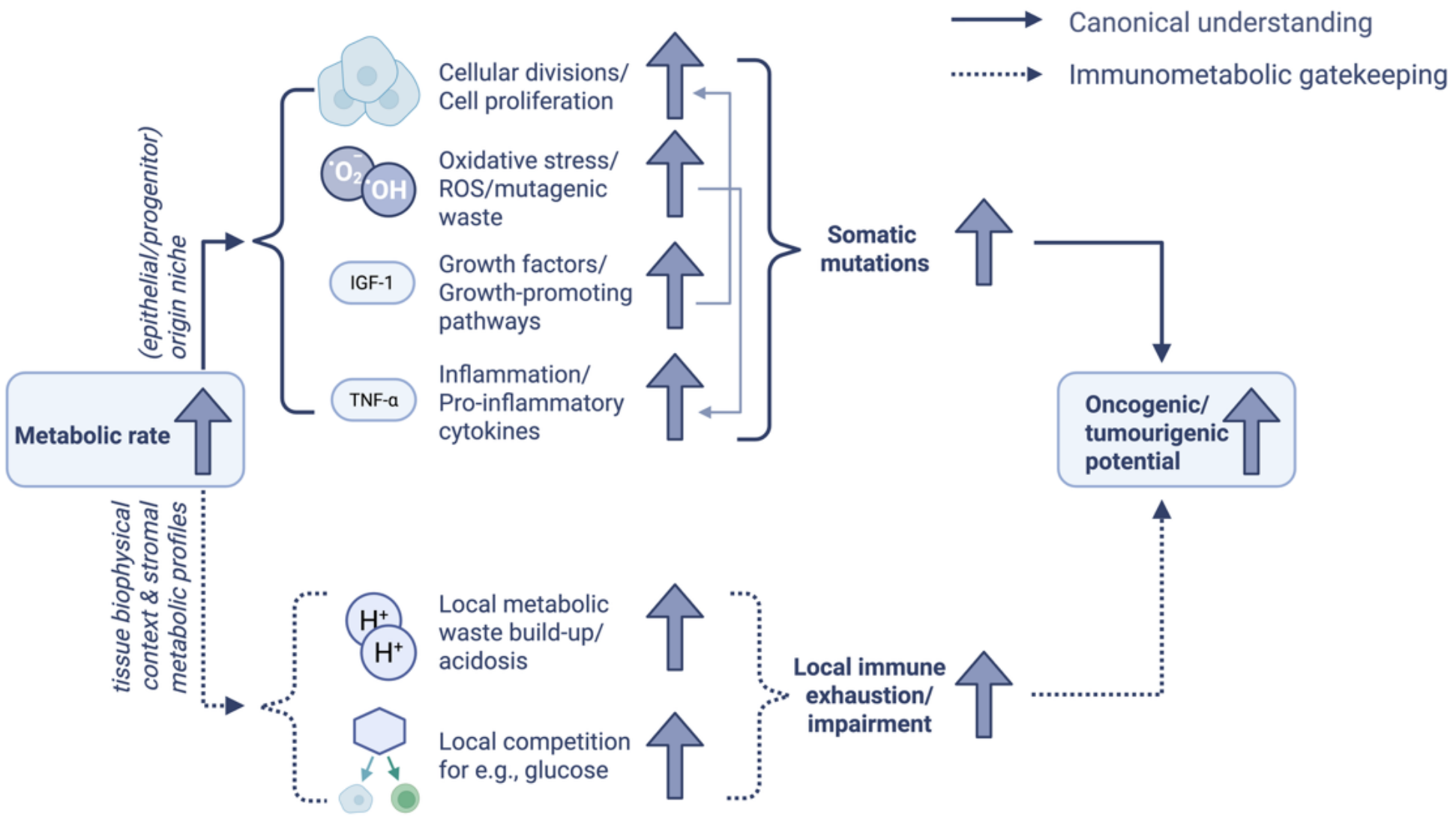

**Figure 1. Schematic overview showing how the immunometabolic gatekeeping framework may complement canonical understanding of the link between high metabolisms and oncogenesis.** While it has been reported how the accumulation of metabolic intermediates and byproducts (e.g., fumarate, 2-hydroxyglutarate, fatty acid derivatives[32,33]) can contribute to tumourigenesis, we here we emphasise a complementary mechanism in which immune exhaustion arises from the local physical and metabolic properties of the tissue.
Note that ROS can increase protein unfolding and proteotoxic stress[34]; the resulting buildup of misfolded proteins and aggregates generates additional waste, further taxing local clearance and promoting immune dysfunction. Collectively, metabolic and proteotoxic debris contribute to tissue environments that impair infiltrating immune cells[28] and facilitate oncogenic/tumourigenic potential.
This diagram was created using BioRender.

**The Warburg Effect as a metabolic barrier to immunity**

While aerobic glycolysis is a normal feature of many metabolically demanding or proliferating cells, it becomes markedly amplified and dysregulated in cancer, where it is co-opted as a dominant metabolic program: the Warburg effect. This shift represents a pathological extension of physiological glycolysis, amplifying metabolic inefficiency and the generation of metabolic “debris”. Compared with OXPHOS, aerobic glycolysis yields far less ATP per unit of glucose but produces a larger flux of byproducts such as lactate and protons. Speficially, when lactate is secreted out of the cell via monocarboxylate transporters[35], it is accompanied by H+, acidifying its extracellular environment if diffusion is slower than secretion (as is expected in solid tissue environments as opposed to e.g., blood).

Whether driven by biosynthetic demand[36,37] or mitochondrial bottlenecks[38–41], this inefficient metabolism converts nutrient flux into waste flux. Lactate and/or associated acidosis suppress nearly every arm of antitumour immunity[42]; impairing T cell motility/infiltration[29,30] and cytokine production[43–45], promoting T-cell exhaustion and PD-1 expression[31,46], skewing macrophage polarisation toward M2-like phenotypes[47–49], and dampening dendritic- and NK cell activity/cytotoxicity[50–52]. In addition, tumour and/or stromal cells can outcompete T cells for glucose via overexpression of GLUT1, reducing T cell glycolysis and IFN-γ production[53]. As such, the Warburg effect becomes not only a metabolic hallmark of cancer but a generator of immune-suppressive environments. Accordingly, systemic (serum) Lactate Dehydrogenase (LDH) levels correlate with inferior outcomes to immune checkpoint blockade across several cancers, and specifically LDH isoform LDHA expression associated with tumour growth, maintenance and tumourigenesis[54]; while often interpreted as a proxy for tumour load, LDH(A) may also capture metabolic waste/lactate-mediated blunting of antitumor immunity.

The Warburg effect may even reframe classic tumour suppressors like p53 beyond 'genome guardians' to metabolic checkpoints: in cancer cells, p53 is regulated by aerobic glycolysis to mount a 'glycolytic stress response' – not to DNA damage, but to the metabolic imbalance itself[55]. Without functional p53, cells fail to buffer this mismatch, continue proliferating under strain, and accumulate damage[55], potentially accelerating immune suppression. This reinforces the idea that tumour suppressors like p53 mediate not only genomic integrity but also metabolic homeostasis, bridging energy stress and immune evasion.

Oncogenic viruses induce many of the same immunometabolic shifts seen in cancer, including (upregulated) aerobic glycolysis, altered mitochondrial and lipid metabolism, and enhanced glutaminolysis and pentose phosphate pathway activity[56–58] – and may even induce fibrosis[56] or deploy several non-cell autonomous mechanisms to reshape the metabolic milieu in the local microenvironment and its constituent stromal and immune cells[59]. These changes support viral replication but also generate immunosuppressive byproducts, promoting local immune dysfunction and facilitating both immune evasion and cellular transformation. Thus, virally driven oncogenesis highlights metabolic reprogramming as a central gatekeeper linking tissue metabolic context to immune competence.

Having considered how tumour metabolism can create an immunosuppressive environment, it is equally important to recognise that immune cells are also metabolically active participants in this ecosystem. For instance, immune cells themselves (especially activated T cells) also shift toward glycolysis during effector activity[60]. This potentially reinforces acidification and lactate enrichment (as detrimental "public goods") within the shared environment where tumour and immune cells reside and co-evolve, leading to a local positive feedback loop that further promotes immune exhaustion – thus potentially explaining how T cell infiltration can be negatively associated with prognosis in high-metabolism tissues. Recent mechanistic work supports this loop: $CD8^+$ T cells rendered metabolically inflexible by mitochondrial PTPMT1 deletion exhibited elevated basal extracellular acidification rates and accelerated exhaustion[61], indicating that metabolically stressed T cells can themselves exacerbate – and be impaired by – microenvironmental acidification. Hence, even if immune cells infiltrate, their own metabolic needs can become maladaptive in an already acidic/lactate-rich niche. Beyond nutrient competition and acidosis, tumours can actively reprogram T cell bioenergetics. A recent study showed that cancer cells transfer mitochondria carrying mutant mtDNA to tumour-infiltrating T cells, where these organelles resist mitophagy and induce metabolic defects and senescence, thereby blunting antitumour immunity and reducing response to immune checkpoint blockade[62]. Such mitochondrial dysfunction would be expected to impair OXPHOS and force a greater reliance on glycolysis, potentially amplifying local lactate production and acidosis. Thus, metabolic competition and feedback between tumour and immune compartments can entrench immune dysfunction, converting local metabolic stress into a self-reinforcing exhaustion loop.

**Age-related and hereditary modulation of immunometabolic priming**

In addition, age-related immune decline further compounds these metabolic effects. With aging, both adaptive and innate immune cells exhibit reduced metabolic flexibility[63–66]. Recent multi-organ proteomic analyses confirm that aging lymphoid tissues, like the spleen and lymph nodes, undergo particularly pronounced declines in protein synthesis, folding capacity, and mitochondrial function[67]. This impairs immune cells' ability to activate, infiltrate, and kill tumour cells in nutrient-competitive or acidified environments. If solid tumours arise in metabolically demanding tissues that are predisposing immunosuppressive TMEs, the aging immune system may be doubly disadvantaged: less capable of metabolic adaptation, and more vulnerable to local inhibitory cues such as lactate or acidosis. This aligns with clinical observations that solid tumour incidence and progression correlate strongly with age, and suggests that while tumour genetics remain important, metabolic compatibility and competition between immune cells and their environment may represent a significant, complementary constraint on immune efficacy (**Fig. 2**). Age-dependent metabolic dynamics may also reflect tropism in paediatric cancers (**Fig. 2**), suggesting that intrinsic metabolic architecture, whether developmental or hereditary, can dictate where tumours preferentially arise. Although germline mutations underlie many paediatric cancers, these do not fully account for tissue tropism; rather, such mutations may require metabolically or developmentally permissive contexts to drive transformation. It is important to note that the same metabolically intense tissues tend to host different cancers across the lifespan – for example, retinoblastoma in the eyes of infants versus uveal melanoma in the eyes of adults – implying that while tissue tropism is conserved, the nature of malignancy reflects the age- and development-dependent interplay between germline predisposition, immune surveillance, and metabolism within that tissue.

This framework also helps explain lesion tropism in certain hereditary cancer syndromes. In VHL disease, for example, tumour formation is not random but concentrated in metabolically intense tissues (e.g., kidney, brain, eye, adrenal gland), which are already primed for high metabolic flux and byproduct stress. This suggests that VHL loss confers a selective advantage primarily in tissues that are highly metabolically active and glucose-permissive. While germline VHL loss impacts every cell in the body, the upregulated glycolysis and suppressed OXPHOS triggered after a second hit may have disproportionately greater consequences in metabolically intense tissues – further limiting glucose availability to local immune cells and intensifying lactate/acid stress. These amplified constraints may predispose to early immune dysfunction and thereby facilitate lesion formation. A similar pattern appears across other hereditary metabolic tumour syndromes. In SDH-deficient syndromes, succinate accumulation drives pseudohypoxia and predisposes tumour formation in select high-metabolism tissues (adrenal gland, kidney, nervous system, stomach, and pituitary gland in brain[68]). Despite distinct molecular drivers, these syndromes share a key feature: tumour formation is not diffuse but restricted to tissues whose intrinsic metabolic architecture can accommodate – or amplify – the metabolic consequences of the germline mutation. This highlights that germline oncogenic mutations operate within pre-existing metabolic niches rather than overriding tissue-level metabolic constraints.

Taken together, we propose that intrinsically high-metabolism tissues are primed to form immune-hostile niches even before transformation. After transformation, the same baseline features (i.e., high nutrient flux, limited byproduct clearance) magnify the immunosuppressive fallout of common metabolic rewiring (e.g., aerobic glycolysis).

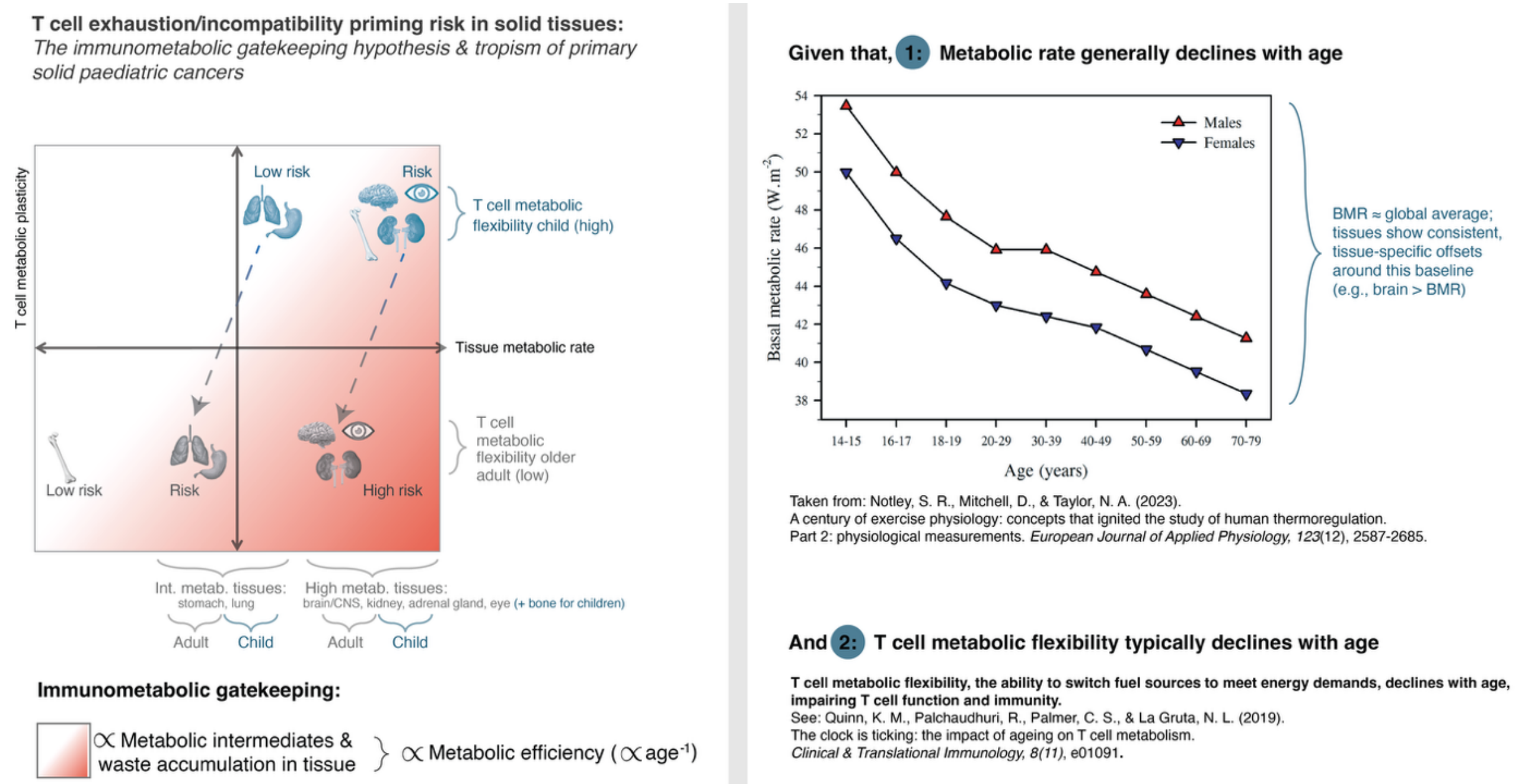

**Figure 2. Paediatric cancer tropism reflects tissue metabolic priming**. Paediatric solid tumours, while rare, arise disproportionately in metabolically intense organs: for instance the brain/CNS (glioma and medulloblastoma), adrenal gland (neuroblastoma), kidney (renal tumours, Wilms), eye (retinal blastoma)[69,70] – consistent with the immunometabolic gatekeeping framework. Primary bone tumours (osteosarcoma, Ewing sarcoma) similarly arise during periods of rapid skeletal growth, when bone and surrounding soft tissues are highly metabolically active, whereas these tissues become metabolically quiescent after growth completes.

Note that while T cell metabolic flexibility generally declines with age[64], neonates/infants may form a unique category: e.g., after infection, neonatal $CD8^+$ T cells upshift glycolysis more than adult cells and form memory poorly unless glycolysis is restrained[71]. Flexibility may thus improve into early childhood, concordant with a developmental shift in T cell differentiation around ~3 years of age[72], and then decline over adulthood. Physical activity mitigates this decline in adults, with older active adults' T cells showing better mitochondrial-glycolytic switching, a higher mitochondrial dependence and lower glucose dependence at rest, as well as a reduced inflammatory phenotype and lower metabolic demand compared to inactive/less-active peers[73].

## Metabolic rate as an evolutionary variable in immune surveillance

This metabolic-immunological perspective may also extend Peto's paradox beyond the genetic basis for tumour suppression. Peto's paradox observes that larger animals with more cells (e.g., whales, elephants) do not have proportionally higher cancer incidence. Traditionally, this is attributed to slower cell division, enhanced tumour suppressor gene copies (e.g., TP53 in elephants), enhanced DNA repair mechanisms (e.g., non-homologous end joining in whales), or lower oxidative damage, all leading to an overall lower mutational burden.

Yet, complementary and more universal explanation may lie in lower species-wide metabolic rates[57]. Across species, metabolic rate scales inversely with body mass: small mammals such as mice exhibit up to sevenfold higher mass-specific metabolic rates and greater spontaneous tumour incidence than humans[75,76], whereas large animals like elephants have markedly lower rates, consistent with allometric scaling laws (e.g., Kleiber's law). As a result, the per-cell or per-gram rate of metabolic byproduct generation – including ROS and lactate – is markedly reduced. The physical process of passive diffusion of small molecules (governed by Fick's laws) remains virtually constant across species and does not scale with metabolic rate. Thus, improved metabolic homeostasis may help explain why large, long-lived species show lower spontaneous tumourigenesis despite greater cellularity and longevity.

Within the human species, the immunometabolic gatekeeping framework appears to hold as well; for instance, men typically have higher basal metabolic rates across tissues than women (**Fig. 2**) and a consistently higher incidence of solid tumours, despite similar environmental exposures[43]. In the example of ccRCC, men are nearly twice as likely to get this cancer than women, with "*the etiology for these disparities not known"* [78]*.* Yet, consistent with our framework, metabolic rate and lactate production flux were shown to be significantly higher for males than for females (in healthy rat kidneys, **Fig. 3**)[79]. Interestingly, bicarbonate levels are not significantly different[79], implying that the buffering for the protons associated with lactate secretion may not be equally accounted for, and thus that acidosis may be higher in male kidneys as well. Moreover, a recent study showcased that top genes that become sex-biased in their expression in kidneys of sexually mature rats are virtually all related to metabolism[80]. Similarly, a recent study in humans found that sex differences in kidney metabolism may reflect sex-dependent outcomes in human diabetic kidney disease[81], further emphasising a fundamental metabolic dimorphism between males and females that likely contributes to differential risks for kidney disease and solid tumours.

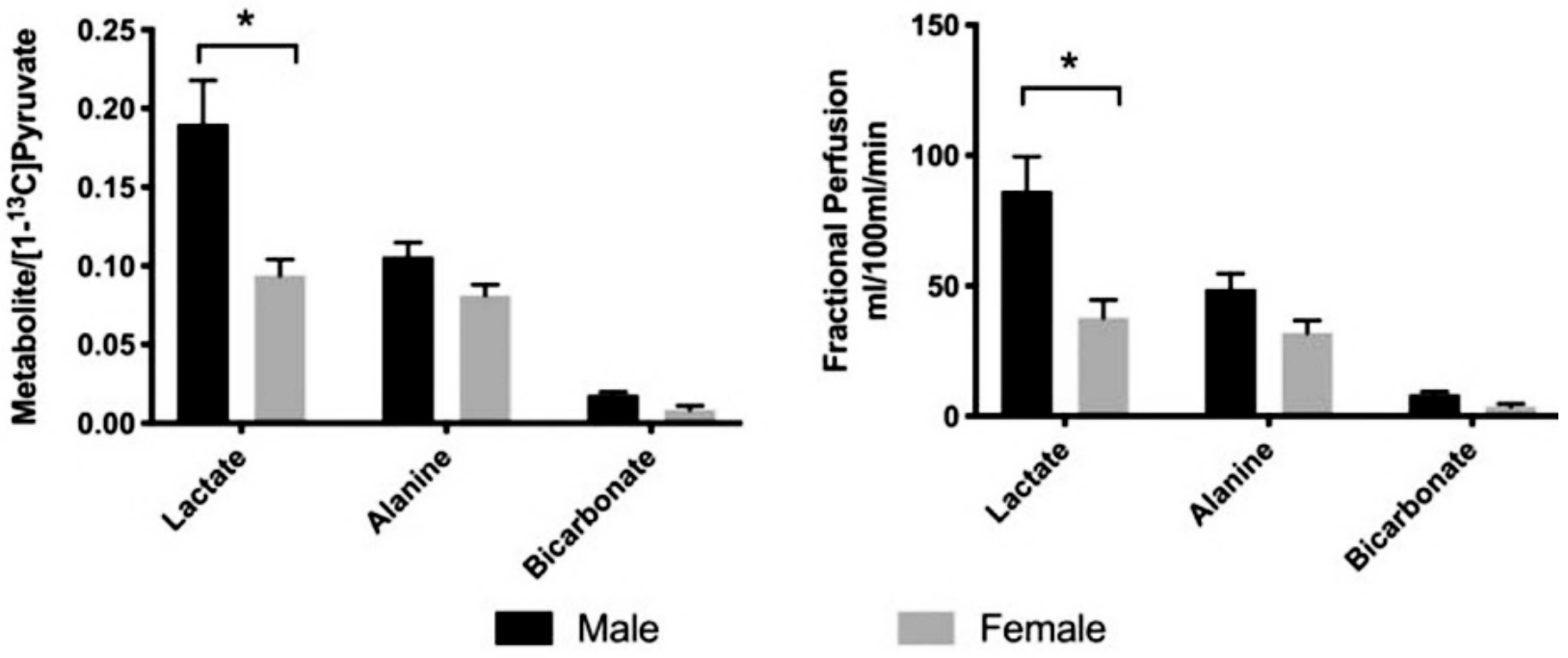

**Figure 3. Metabolic dimorphism in renal lactate levels parallels cancer susceptibility.** Sex-specific levels of lactate measured in healthy kidney tissue of rats, from [79]. Regardless of metabolite quantification/normalisation method, lactate production for males appears double that of females.

Furthermore, intra-tissue metabolic heterogeneity can shape both the site of tumour origin and the relationship between immune infiltration and prognosis. Tumours arising from metabolically intense and/or inefficient (i.e., glycolytic) compartments may be more prone to waste accumulation, local acidosis, and immune exclusion, leading to weaker correlations between T cell infiltration and outcome. In the kidney, for example, papillary and particularly clear cell renal cell carcinomas, which originate from highly oxidative, gluceogenic and glycolytically flexible proximal tubule (PT) cells, exhibit immune infiltration that is ineffective or even prognostically adverse[8, 82–87]. By contrast, chromophobe RCC, derived from the less glycolytic intercalated cells of the distal nephron, tends to show a more favourable infiltration-prognosis relationship[9,82,85,88,89]. Importantly, although these subtypes originate from niches situated within the metabolically active renal cortex, their distinct metabolic programmes[82] appear to determine whether immune infiltration remains functionally productive, linking metabolic origin to immune competence and clinical outcome.

**Metabolic sanctuaries: primary tumour rarity at the metabolic extremes**

Primary cancers arising from tissues at two metabolic extremes: highly oxidative heart tissue and metabolically quiescent fat (i.e., adipose) tissue are exceedingly rare[90–92]. This rarity is typically attributed to the fact that adipocytes and cardiomyocytes are fully differentiated, non-proliferative/low-regenerative, and genetically stable, rendering them less prone to mutation accumulation[93]. However, this raises an important contrast with other tissues that also contain differentiated, low-turnover cells, such as renal tubule epithelial cells, which do frequently give rise to cancers (i.e., ccRCC). Moreover, linking low-proliferation to a "fewer mutations" explanation is no longer sufficient: modern sequencing shows that somatic mutations and mosaicism accumulate even in low-proliferative or post-mitotic tissues. Instead, it may be the unique metabolic ecology of these tissues, i.e., how they process energy and neutralise its byproducts, that grants them exceptional resistance to transformation.

Indeed, these "metabolic sanctuaries" illustrate how both high flux with efficient waste clearance and pH homeostasis, and low flux with limited biosynthetic drive, can protect against transformation. In brown adipose tissue, uncoupled respiration dissipates energy as heat rather than ATP/biomass[94], producing minimal metabolic waste despite high oxygen use when active[94,95]. Interestingly, cold-induced brown fat activation was shown to impede glycolysis and reduce tumour growth in multiple cancers in murine models[96,97]. In cardiomyocytes (and red skeletal muscle), efficient fatty acid oxidation[98] and lactate consumption[99,100] sustain their unique specialisation on ATP production for contraction[101] while preventing lactate accumulation or acidosis. Cardiomyocytes are further uniquely specialised for pH homeostasis: they express exceptionally high levels of ion transporters and carbonic anhydrases that maintain acid-base balance even under intense metabolic load, buffering both intracellular and extracellular pH[102]. Moreover, any extracellular 'cost' of intracellular pH maintenance (e.g., $CO_2$ production as a consequence of $H^+$ neutralisation[102]) is quickly diluted by the high vascular perfusion of cardiac tissue, facilitating rapid clearance of (downstream) metabolic byproducts.

Both metabolic extremes of cardiomyocytes and adipocytes thus effectively maintain extracellular pH homeostasis and preclude the Warburg-like switch associated with tumourigenesis. In contrast, metabolically demanding and flexible tissues like the renal proximal tubule, capable of toggling between oxidative, glucogenic and glycolytic states, may be uniquely susceptible to local waste accumulation, associated immune exhaustion and ultimate oncogenic reprogramming. Thus, tissues that are either metabolically quiescent or specialised to avoid metabolic/glycolytic byproduct buildup resist the formation of immune-suppressive niches, offering natural metabolic models of tumour resistance. Consistent with this, metastatic tumours to the heart are 100- to 1000-fold more common than primary cardiac tumours[103], suggesting that only evolved cancer cells with additional immune-evasion mechanisms (beyond those mediated by local nutrient competition, waste accumulation, and acidosis) can successfully colonise such metabolically protected tissue.

**Bats and naked mole rats: the metabolic antitheses of cancer?**

The cancer-resistant naked mole-rat (NMR) offers an illuminating case study of species-level metabolic extremes. Living in chronically hypoxic burrows, NMRs exhibit extreme metabolic suppression – up to 85% in acute hypoxia – and downregulate glycolysis, β-oxidation, and ATP-consuming processes[104]. Unlike most mammals, they do not respond to hypoxia with increased glycolysis, thus producing minimal lactate. As stated by Farhat et al., "*Hypoxic NMRs can afford to slow down glycolysis because they rely on the suppression of aerobic metabolism that likely spares small carbohydrate stores and minimizes the accumulation of anaerobic end products*" [104]. Moreover, Hadi et al., showed that the NMR's unique TME is what stops the initial stages of cancer from developing into tumours, rather than a cancer resistance mechanism intrinsic to NMR somatic cells as previously thought[105].

Bats are another group of small mammals known for their exceptional resistance to cancer[106]. Despite having the highest metabolic demands of any mammal due to the energy-intensive nature of flight (with energy expenditures 3 to 5 times greater than those of mice[107]), bats rarely develop cancer and live significantly longer than expected. Bats make up 18 out of 19 size-corrected mammalian species with natural lifespans longer than our medically-assisted ones (1/19 is the NMR)[108]. While this resistance has mostly been attributed to adaptations that help bats tolerate DNA damage from elevated metabolic stress[108], a recent study points to a distinct metabolic mechanism via the downregulation of three key genes – HIF1A, COPS5, and RPS3[109]. The downregulation of HIF1A (and secondarily COPS5) dampens hypoxia signalling, leading to reduced angiogenesis, aerobic glycolysis and lactate production, contributing to the species' reduced cancer susceptibility.

This suggests that these exceptional mammals' resistance to cancer may not just lie in (taxon-specific) genetic or proteostatic mechanisms, conventional areas of focus[105,106,108,110], but also in the overall avoidance of byproduct-rich, acidic, immune-suppressing microenvironments. The NMR's unique metabolism allows survival without accumulation of immunosuppressive metabolic byproducts, offering a natural model of lactate-low hypoxia. Likewise, bats maintain extreme oxidative flux with efficient waste clearance and muted hypoxia signalling, preventing the formation of acidified, immune-suppressive microenvironments despite intense metabolic activity.

Beyond these cancer-resistant species, immunometabolic gatekeeping model holds in broader cross-species comparisons: a comparative genomics approach across nearly 200 vertebrates found that genes whose conservation levels correlate negatively with cancer resistance are *"enriched for metabolic functions"* [111], suggesting that metabolic activity is a key contributor to cancer incidence disparity across species. Notably, our immunometabolic gatekeeping framework interprets this metabolic signal as a reflection of metabolism's deeper impact: as a mediator of immune ecology governing cancer susceptibility across species (**Fig. 1**).

**Desmoplasia as a physical constraints of metabolic reprogramming and waste accumulation**

Beyond cellular metabolism, physical properties of a tissue's microenvironment can shape the local link between metabolic and immune behaviour. Pan-cancer analyses reveal that fibrotic and immune-fibrotic tumours fare worst clinically, pointing to stromal stiffening as a key barrier to anti-cancer immunity[112] – a constraint that may operate even before tumour initiation. Fibrosis and desmoplasia increase extracellular matrix (ECM) stiffness, which restricts diffusion of metabolites, protons, and other debris while directly reprogramming stromal cells toward glycolysis: a study in mammary models has shown that, even in the absence of tumour cells, normal stromal cells cultured on stiff ECM resembling tumour desmoplasia undergo glycolytic reprogramming, including upregulation of GLUT1 and MCT4 and increased lactate production[113]. ECM stiffness was also shown to enhance aerobic glycolysis in mesenchymal stem cells[114]. This demonstrates that ECM stiffness alone – independent of malignant signalling – is sufficient to induce a protumour metabolic state, highlighting the role of the physical microenvironment in priming tissues for malignancy.

This metabolic reciprocity between tumour and stroma thus challenges the assumption that metabolic reprogramming starts with cancer cells. Therefore, locally elevated organic waste may not only be a product of enhanced and/or inefficient metabolic flux but may also reflect physical constraints on clearance within fibrotic or densely stromalised normal (or transformed) tissue, favouring local acidosis and immune exhaustion. Consistent with this, fibrotic diseases across organs precede and predict cancer risk, from cirrhosis to chronic pancreatitis. Generally, this link between fibrosis and cancer risk is attributed to its impact on cellular transformation, cell signalling, and proliferation[115]. Here, we emphasise a potential additional role of fibrosis as a mediator of metabolism, metabolic/proteotoxic waste accumulation and thus local immune exhaustion. ECM stiffening can thereby act as an upstream driver of metabolic inefficiency and immune dysfunction, establishing a physical layer of immunometabolic gatekeeping that links tissue mechanics to tumour susceptibility.

**Geometric constraints on immunometabolic equilibrium**

The immunometabolic gatekeeping framework extends beyond explaining existing paradoxes in cancer immunity to offer a generative lens for understanding how spatial features of tumour growth influence metabolic stress, immune accessibility, and therapeutic efficacy. For instance, a persistent clinical observation in ccRCC is that primary tumours smaller than ~3 cm in diameter rarely metastasise[116], despite some already containing known aggressive genetic drivers. In VHL disease and in sporadic small renal masses (SRMs), this “3 cm rule” is used to clinically guide surveillance and surgical timing. Larger tumours, in contrast, show higher metastatic potential and correlate with greater evolutionary divergence. This paradox has prompted speculation about possible geometric or environmental thresholds that influence this intriguing bifurcation in tumour behaviour unexplained by genetic profiles alone. While this rule is often viewed as a practical cutoff, it may reflect a deeper biophysical constraint: a spatial-metabolic tipping point. Tumours below 3 cm retain high surface-area-to-volume ratios that support efficient diffusion of both oxygen and waste, limit acid build-up, and maintain immune accessibility and equilibrium. This concept is supported by work of Hakimi et al., showing that metabolite accumulation (e.g., dipeptides) increases with clinical stage and size of tumours in VHL patients[32], and explored as a diffusion-reaction model in **Supporting Materials II**.

Within this framework, genetically aggressive tumours may remain indolent while still small because diffusion is sufficient to prevent the build-up of lactate, protons, and other metabolic or proteotoxic byproducts that impair immune function. Moreover, because VHL loss drives constitutive HIF activation and early angiogenesis, smaller ccRCC lesions likely remain relatively well oxygenated, making true hypoxia an incomplete explanation for a size-dependent shift in immune efficacy. Instead, the “3 cm rule” may represent a diffusion-limited transition in immunometabolic equilibrium rather than a discrete genetic or angiogenic event. By linking tumour geometry to waste accumulation and immune suppression, this extension of the immunometabolic gatekeeping framework provides a mechanistic rationale for why ccRCCs below ~3 cm remain non-metastatic despite oncogenic potential. Similar size-dependent transitions in e.g., metastatic competence in other cancers may also reflect immunometabolic tipping points (e.g., Breslow thickness in cutaneous melanoma).

Such geometric constraints operate at the macroscopic scale of tumour architecture, but similar principles may apply within tumours themselves: tumour heterogeneity is often qualified and quantified genetically, yet studies in e.g., NSCLC demonstrate that metabolic heterogeneity exists within tumours as well[117]. Such intra-tumoral metabolic heterogeneity may further shape local immune accessibility, resistance to therapy, and evolution of immune escape mechanisms. Indeed, recent functional analyses of subclonal immune escape at single clone resolution in NSCLC has revealed intra-tumoural variation in immune escape mechanisms[118], suggesting that distinct tumour regions may harbour different degrees of metabolic hostility and immune dysfunction. Incorporating this layer of heterogeneity into the immunometabolic gatekeeping model could improve predictions of treatment response; especially in spatially complex or treatment-resistant tumours.

**Metabolic treatments & interventional studies**

With an immune-metabolic focus, several studies have explored the potential of targeting acidosis or lactate concentrations (i.e., the consequence of high glycolytic rates), or lowering the underlying metabolic rates altogether via e.g., caloric restrictions[23,119] or glucose-low diets[120], on reducing tumour growth/invasion. A recent study integrating human and mouse data showed that modulation of dietary amino acids altered metabolic flux within the tumour, restrained glioblastoma growth, and enhanced standard therapy *in vivo*, providing proof-of-concept for metabolism-informed combination strategies[36]. Metabolic modulators already in clinical use, such as GLP-1 receptor agonists (which lower systemic glucose and insulin), have shown early signals of anti-tumour benefit in preclinical and epidemiologic studies[121,122], further supporting host-targeted combination strategies. Indeed, such host-level shifts may mitigate the metabolic tone that fosters immunosuppressive TME, positioning GLP-1 agonists as systemic 'immunometabolic normalisers', and thus acting upstream of tumour metabolism itself (supported by evidence that GLP-1 therapy in people with obesity restores immune metabolism and effector function[123]). Preclinical models have shown that oral administration of pH-buffer sodium bicarbonate was sufficient to increase peritumoral pH and inhibit tumour growth and local invasion[124]. Others have shown how (selectively[125]) targeting key proteins in metabolism, via e.g., GLUT1 inhibitors[126,127], CD36 inhibitors[128], Monocarboxylate Transporter (MCT) inhibitors[129,130] and Lactate Dehydrogenase A (LDHA) inhibitors[131–135], can suppress tumour growth and may enhance the efficacy of conventional therapies.

However, caution in designing these treatment approaches is warranted. For instance, Apostolova & Pearce[35] argue against therapies aimed at lowering lactate concentrations, since a proportion of lactate is utilised as a metabolic fuel by immune cells and/or healthy tissue. While metabolic therapies or caloric deprivation/fasting have yielded promising results in e.g., potentiating the effects of chemo- and radiotherapy, tyrosine kinase inhibitors, immunotherapy, and hormone therapy[136,137], they may have serious side effects (e.g., weight loss[137]) and may be particularly challenging for e.g., patients recovering from chemotherapy[138]. The purpose of this work is not to review existing interventions in the space of immunometabolism (as done by e.g., [54,139–142]), but rather to present a unifying framework reconciling key paradoxes within cancer biology.

**An emerging immunometabolic paradigm**

Together, our findings outline an "immunometabolic gatekeeping" model in which the metabolic properties of healthy tissues shape immune competence before and after tumour initiation. Tissues with high metabolic activity and limited capacity for waste clearance are predisposed to local accumulation of lactate, protons, ROS, and proteotoxic byproducts. These features create niches in which infiltrating immune cells, particularly T cells, lose metabolic plasticity, exhaust more rapidly, and fail to control early neoplastic lesions. In such settings, high T cell infiltration may not predict favourable outcomes; instead, it may simply mark immune cells entering an environment they cannot functionally withstand.

This framework helps reconcile multiple longstanding paradoxes across cancer biology. Linking Peto's paradox (lower cancer incidence in large, low-mass-specific-metabolism species), tissue-level discrepancies in T cell prognostic value, and size-dependent transitions such as the "3 cm rule" in ccRCC, to a common variable: the ability of tissues to buffer metabolic waste and maintain immune-permissive conditions. High-flux tissues such as the kidney, brain, and eye are thus primed to form acidic and/or waste-rich microenvironments both before and after transformation, whereas metabolically quiescent tissues (e.g. adipose) or highly oxidative but efficiently buffered tissues (e.g. heart, skeletal muscle) rarely give rise to primary cancers. Although the eye and brain are immune-privileged tissues, their intrinsically high metabolic activity makes them susceptible to waste accumulation regardless of immune privilege. Long-lived small mammals, namely bats and naked mole rats, illustrate the same principle at species scale: both avoid byproduct-rich and/or hypoxia-driven metabolic states despite extreme or chronic energetic demands, thereby preventing the formation of immune-suppressive microenvironments.

By reframing cancer vulnerability as a property of tissue-intrinsic metabolic licensing rather than solely tumour-intrinsic mutations, this model complements classical genetic frameworks (e.g., Tomasetti & Vogelstein[143]), and tumour-stroma co-evolution models (Hanahan & Weinberg[144]). It suggests that immune efficacy is not a fixed property of the immune system but is gated by the metabolic terrain in which immune cells operate. This perspective expands the explanatory reach of existing cancer theory and supports the development of immunometabolic interventions that act on the host environment, not just on tumour cells.

The framework also generates translational implications. First, tissues such as the kidney should be viewed not as passive sites of tumourigenesis but as metabolically primed environments in which early lesions may rapidly acquire immunological privilege. Because these metabolic profiles are physiologically essential (e.g. high flux in neurons), interventions should aim not to reprogram baseline tissue function but to buffer early metabolic shifts such as local acidification or GLUT1 upregulation, particularly in genetically or clinically high-risk individuals.

Second, host-level metabolic states such as obesity, diabetes, or elevated basal metabolic rates likely intensify these tissue-level vulnerabilities. Hyperglycaemia increases substrate availability for glycolysis; systemic inflammation or insulin resistance reduces immune metabolic fitness. These factors may help explain epidemiological links between metabolic syndromes and increased cancer incidence and/or poorer outcomes. Additionally, recent increases in early-onset solid cancers may also reflect changes in host-level metabolic fitness: sedentary lifestyle and metabolic syndrome impair T cell mitochondrial flexibility and resilience to acidic or nutrient-competitive environments[73], potentially lowering the immunometabolic threshold for tumour initiation and/or growth across tissues.

Third, the model provides a mechanistic rationale for metabolic-immune combinatorial therapy. Normalising the metabolic hostility of the TME (via buffering agents, GLUT1 inhibition, lactate/MCT targeting, or systemic metabolic modulators) may reopen access to tumour sites but may be insufficient alone if infiltrating T cells are already exhausted or epigenetically fixed. In such cases, metabolic interventions may synergise with immune checkpoint therapy by reducing environmental suppression while reinvigorating the T cell compartment. Early studies support this approach[125,145–147], though future trials may need to stratify patients by tissue-level and host-level metabolic compatibility, not only by canonical tumour genomics.

Importantly, immunometabolic gatekeeping is not proposed as a single determinant of cancer incidence. Tissue tropism in paediatric solid tumours and VHL disease, while consistent with the model, also reflects developmental context, progenitor pools, stem cell dynamics, and mutational processes. Spatial and temporal heterogeneity within tumours adds further layers: metabolic gradients, diffusion barriers, fibrosis, and local myeloid programming can vary regionally, shaping transient windows of immune accessibility. Systemic states – childhood metabolic rates, sex-specific metabolism, germline predispositions, hyperglycaemia, chronic inflammation – modulate these local conditions rather than replace them. Immunometabolic gatekeeping therefore represents an upstream constraint that interacts with, rather than overrides, classical genetic and immunological determinants of tumour risk.

Taken together, this framework recasts cancer development as a dynamic interplay between tissue metabolism and immune cell state. It suggests new principles for A) identifying individuals and tissues at risk, for B) predicting when immune surveillance will fail, and for C) designing interventions that target the metabolic terrain in which tumours emerge. By shifting focus from the cancer genome to the ecological conditions that license or restrict immune function, immunometabolic gatekeeping offers a coherent lens through which to integrate diverse observations across tumour biology, species biology, and clinical oncology.

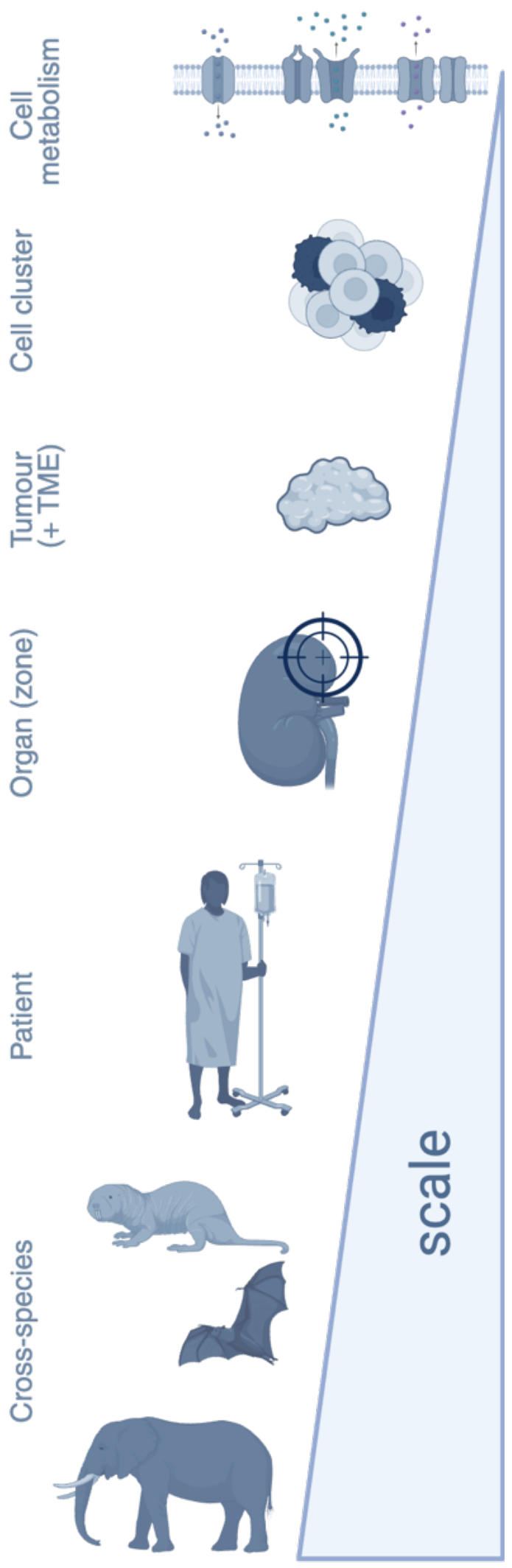

| Component | Description | Key support |
|---|---|---|
| **Healthy cell metabolic rate & nutrient-permissiveness** | The metabolic rate of a differentiated, tissue-resident cell is shaped by the metabolic demands of its resident tissue and constrained by its epigenetically encoded, though partially flexible, phenotype. This may prime the cell for metabolic adaptability post-transformation and/or cancer-associated reprogramming | **Table 1** |
| **Waste accumulation, nutrient-competition & acidosis exhaust infiltrated T cells** | Consequences of (fast) metabolism in restricted space; creation of immunosuppressive niches where metabolic byproduct/protein aggregates/stress granules generation > diffusion, impairing T cell function even if infiltration is present | **Table 1**<br>[25], [148], [28] |
| **Microbial modulation** | Oncogenic viruses reprogramme cellular metabolism pre-transformation. Certain bacteria can enhance Warburg metabolism and lactate buildup, further shaping immune-hostile microenvironments | [149], [150], [151] |
| **Spatially- or temporally consistent constraints on immune efficacy** | • *Spatial:* Tumour size affects SA:V ratio; small tumours allow better metabolic waste diffusion, preserving immune access and equilibrium (e.g., 3 cm rule in VHL-disease & ccRCC)*<br>• *Temporal:* Circadian modulation of glucose/lactate transport affects daily fluctuations in TME permissiveness and may influence (time-of-day dependent) immune (therapy) efficacy* | *Spatial:* [32]<br>*Temporal:* [152], [153] |
| **Tissue zonation & tropism** | • *Zonation:* Tumour origin often maps to metabolically intense intra-tissue regions, which are more prone to waste accumulation and thus forming immune-exhausting niches<br>• *Tropisms:* High metabolic tissues align with VHL lesion and paediatric solid cancer tropisms | Ductal cells PDAC, proximal tubule in ccRCC, perivenous zone in HCC<br>*Tropisms:* **Figure 2** |
| **Patient-level factors influencing metabolic 'tone'** | Individual characteristics that elevate systemic or local metabolism and waste production, predisposing to metabolically immunosuppressive environments. These factors may increase cancer risk or worsen prognosis of solid tumours. E.g.,:<br>• *Systemic hyperglycaemia* increases glucose availability, fuelling lactate production and fostering immune cell exhaustion;<br>• *Sex differences***:** male basal metabolic rates and waste production are higher than that of females at any given age, aligning with higher incidence of certain solid tumours like ccRCC in males | Disorders w/ hyper-glycemia (i.e., T2D, obesity) are associated with higher risk for many solid tumours [154], [155]<br>Lactate in healthy kidneys twice as high in males [79] |
| **Long-lived small mammals** | Long-lived small mammals like bats and NMRs avoid metabolic byproduct accumulations and/or Warburg effects via suppressed (glycolytic) metabolisms or altered hypoxia responses | Downregulation of aerobic glycolysis in NMR[105] & bats[109] |
| **Peto's paradox** | Large animals with more cells avoid metabolic waste accumulation via lower mass-specific metabolic rates (i.e., basic allometry) | Metabolic rates scale with body size while passive diffusion of organic waste does not |

**Figure 5: Immunometabolic gatekeeping framework summarised.** Asterices indicate follow-up hypotheses.
This diagram was created using BioRender.

**Acknowledgements**
N.I.v.d.B and K.M. are funded by by the MANIFEST programme, core funded by the UK Government's Office for Life Sciences and the Medical Research Council (Grant Ref: MR/Z505158/1).
M.E. is funded by the Cancer Research UK Convergence Science Centre PhD Programme (CANCTA-2024/10007).
O.B. is funded by the Royal Marsden Cancer Charity (1095197).
O.S.A. is funded by an NIHR Clinical Lectureship in Renal Medicine.
X.F. is funded by the McNab bequest at the Cancer Research UK Scotland Institute and receives further support from the MANIFEST programme (Grant Ref: MR/Z505158/1).
S.T. is supported by Cancer Research UK (A29911, A27412 for Cancer Research UK Manchester Institute), the Francis Crick Institute (core funding from CRUK FC10988, UK Medical Research Council FC10988, and the Wellcome Trust FC10988), the NIHR Biomedical Research Centre at the Royal Marsden Hospital and Institute of Cancer Research (A109), the Royal Marsden Cancer Charity, the Rosetrees Trust (A2204), Ventana Medical Systems Inc. (10467, 10530), the US National Institutes of Health (U01 CA247439), the Melanoma Research Alliance (686061), the US Department of Defense (W81XWH-22-1-0652), the VHL Alliance (PRJ_20450), and the UK Office of Life Sciences and Medical Research Council (MR/Z505158/1).

We further thank Anne-Laure Cattin, Matthew Vander Heiden, Karen Vousden, Husayn Pallikonda and Ilaria Malanchi for their helpful insights in discussions about the hypothesis. We thank Andrew Porter for his useful feedback on the manuscript.
We also thank the Wellcome Cancer Evolution summer school and Cambridge University Isaac Newton Institute's Mathematical Foundations of Oncological Digital Twins workshop (OOEW07) for helpful insights and discussions driving the further development of the hypothesis.

**Author contributions**
N.I.v.d.B conceived of the hypothesis, performed the data analyses, and wrote the first draft of the manuscript. S.T. and X.F aided in further developing the hypothesis, provided ideas for analyses and co-wrote the manuscript. M.E. aided with developing the hypothesis. M.E., K.M., O.B. and O.S.A. provided feedback on the manuscript.

**Disclosure statement/competing interests**
Samra Turajlic has received speaking fees from Roche, AstraZeneca, Novartis, and Ipsen, and has filed patents on Indel mutations as a therapeutic target and predictive biomarker (PCTGB2018/051893) and Clear Cell Renal Cell Carcinoma Biomarkers (P113326GB).

**Data & code availability**
All scripts used in this study to analyse publically available data are available at https://github.com/NaomiIrisvdBerg/immunometabolic_gatekeeping.
The repository includes the full data processing pipeline and figure generation scripts supporting the findings of this manuscript. Processed datasets are provided; raw public datasets are referenced within the Methods embedded in the respective Supplementary Materials.

**Supporting materials IA – Single cell gene expression data across normal human tissue (CELLxGENE)**

Human single cell gene expression data were downloaded as a .csv file from the CELLxGENE portal (October 2025). The dataset includes expression metrics for various genes across multiple tissues and cell types, including:

i. Tissue and cell type annotations
ii. Gene symbol
iii. Log-normalised mean expression (Expression)
iv. Scaled expression z-scores (Expression, Scaled)
v. Cell count per cell type and tissue
vi. Number of cells expressing a given gene within the respective cell type
vii. Publication/source
viii. Sex

The dataset was read and processed in R (v 4.5.1). The only cancer type listed in **Table 1** for which we did not find a straightforward healthy (i.e., 'normal') tissue baseline was head and neck cancer. While CELLxGENE does have single cell expression data for tissues such as 'Nose' and 'Tongue', calculating composite expression profiles from these tissues was considered methodologically unrepresentative, and thus not included in this analysis.

To examine baseline metabolic versatility in healthy tissue, we focused on fibroblasts and endothelial cells; stromal cell types universally present across tissues and important non-immune cell types in tumour TMEs[156]. As expanded upon in the main text, fibroblasts are often reprogrammed into cancer-associated fibroblasts (CAFs), which actively contribute to the metabolic rewiring and acidification of the tumour microenvironment. Understanding baseline fibroblast and endothelial cell metabolic potential gives insight into how readily a tissue could form an immunosuppressive stromal niche before and/or after transformation.

For any unique combination of Tissue, Cell Type and Gene, the `Number of Cells Expressing Genes` had to be $\geq 10$ to meet the QC threshold. Across Cell Type == "fibroblast" | "endothelial cell", gene expression was weighted by the cell type's cell count within each tissue, yielding an average expression per gene that reflects the abundance of each intrinsic cell type in the tissue. All 'focus' tissues (i.e., solid tissues listed in **Table 1**, in accordance with [10]) were represented by ≥5 publications, except for the bladder. Weighted expressions were therefore calculated across 'bladder organ' (2 publications) and 'urinary bladder' (1 publication), to arrive at a total of 3 publications representing single cell expression profiles of normal 'bladder'.

Similarly, T cell exhaustion was gauged by "TOX" expression for Cell Type == "T cell", and subjected to similar data processing as conducted for gauging relative SLC2A1 expression across stromal cells as outlined above (without the need to weight expression across multiple cell types as for this gene we only investigated one cell type).

Moreover, to gauge pH/chemical homeostasis, the expression of a panel of genes was compared: "SLC9A1", "SLC4A2", "SLC4A3", "CA2", "CA9", "CA4", "SLC4A7", "SLC4A4", and "SLC4A5". Weighted expressions were normalised per gene to its cross-tissue maximum to allow balanced cross-tissue comparisons within each marker panel. The normalised values were then averaged across the panel to generate a composite pH-homeostasis score for each tissue. Relative stromal SLC2A1 expression, composite pH-homeostasis score, and T-cell TOX expression were then compared across tissues listed in **Table 1** and categorised into the relative bins shown.

**Supporting materials IB – Cross-tissue mitochondrial respiration in normal mice**

We used the data supporting figure 1 (elife-96926-fig1-data2-v1) from [12] to create metabolic intensity classifications reported under the column 'Mitochondrial resp. average (complex IV) in mice' in **Table 1.** Complex IV (cytochrome c oxidase) was used as a proxy for mitochondrial respiratory capacity because it is the terminal oxidase of the electron transport chain.
Tissue labels were collapsed into predefined focus categories (tissues analogous to those listed in **Table 1**) as follows: Kidney (Kid-Cortex and Kid-Med), Brain (Cortex), Eye (Eye), Stomach (Stomach), Pancreas (Pancreas), Lung (Lung), Liver (Liver), Fallopian tube (records with tissue "Test Fallop tube" and sex = Female), Skin (Skin), and Colon (Proximal colon and Distal Colon).

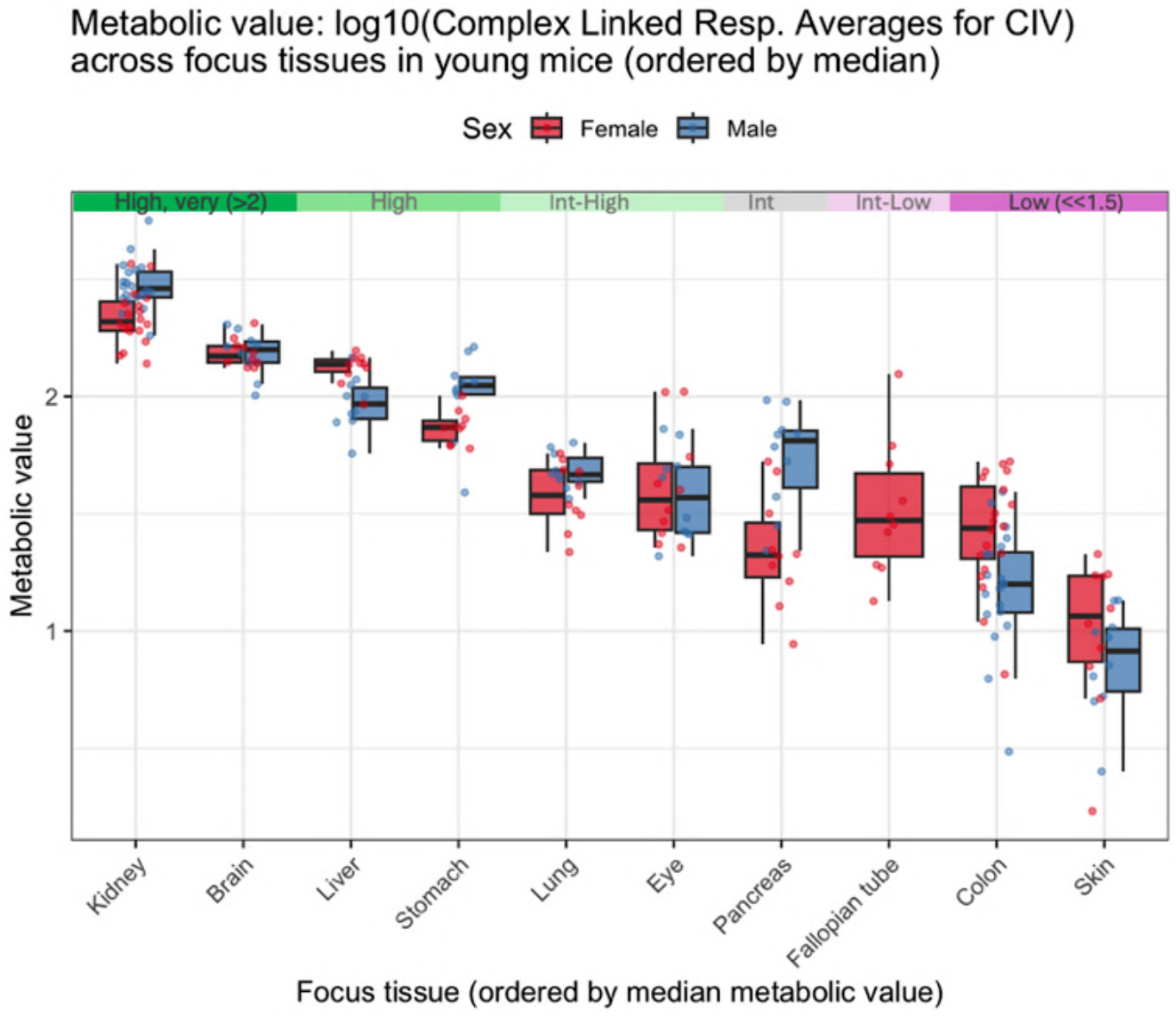

**Suppl. Fig. 1. Ranking of tissues (as found in Table 1) by their complex-linked respiratory averages, CIV.** Boxplots were ordered by the overall (cross-sex) median metabolic value (log10(Complex Linked Resp. Averages for CIV)).

When using the "Complex Linked Resp. Average" value following normalisation against the tissue's mitochondrial content, "Complex Linked Resp. Avg/MTDR", across all complexes measured (CI, CII, CIV), the ranking remained robust (except for Pancreas having a median lower than that of Fallopian tube and Skin):

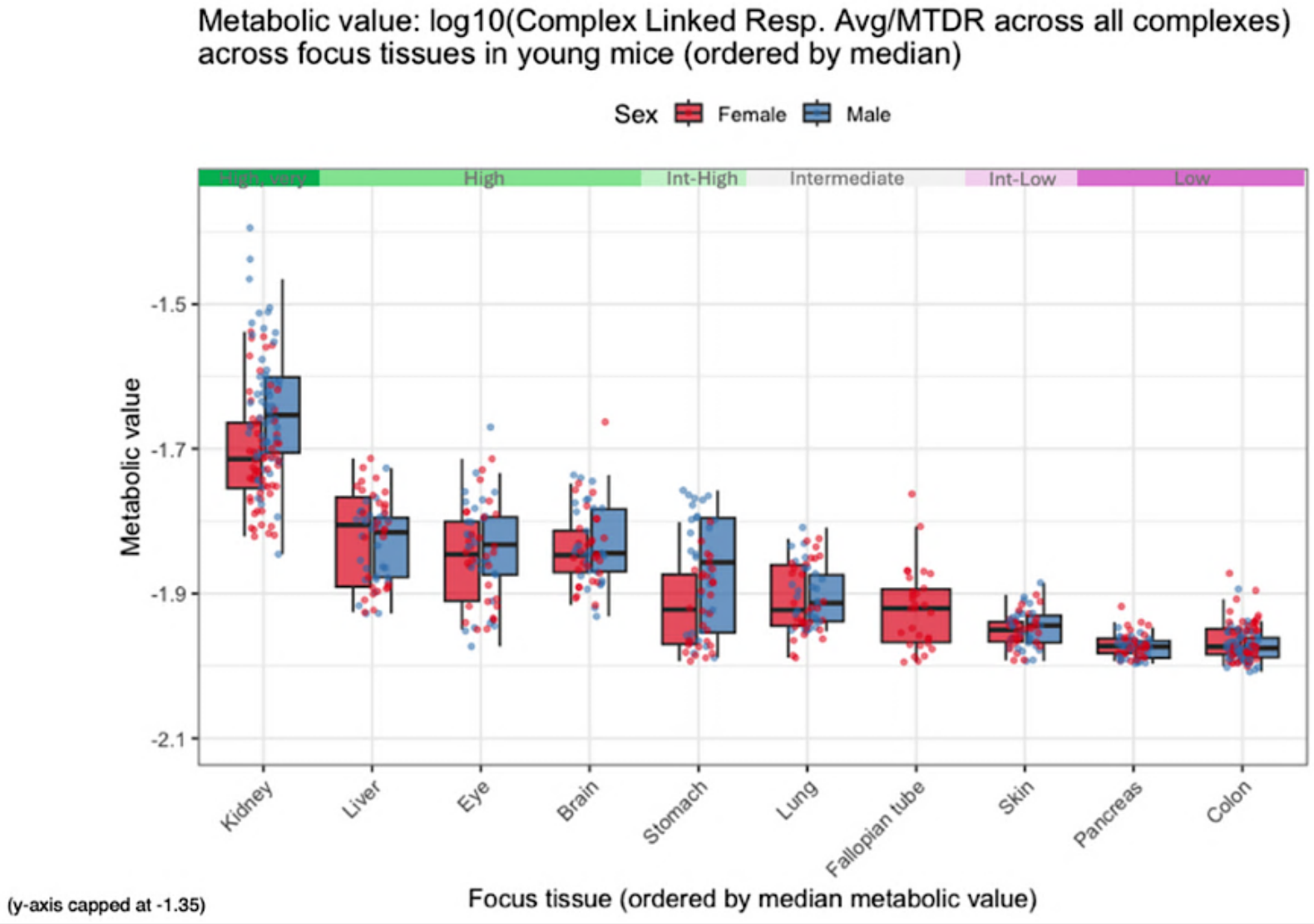

**Suppl. Fig. 2. Ranking of tissues by their complex-linked respiratory averages normalised by mitochondrial density (CI, CII, CIV).**

**Supporting materials II – A minimal diffusion-reaction model illustrating a geometry-driven immunometabolic tipping point at ~3 cm in ccRCC**

**Overview and clinical motivation**

Small (<3 cm) clear cell renal cell carcinomas (ccRCCs) infrequently metastasise, whereas larger tumours show a marked increase in metastatic potential – a phenomenon referred to as the "3 cm rule" [116]. Importantly, this transition typically occurs without acquisition of new driver mutations, suggesting that the shift in metastatic competence reflects microenvironmental changes rather than genetic evolution.

To investigate whether biophysical constraints imposed by tumour geometry may generate such a transition, we constructed a steady-state diffusion-reaction model describing lactate accumulation, oxygen availability, and cytotoxic immune cell infiltration within spherical tumours of varying radii.

**Biological rationale**

Constitutive HIF stabilisation following VHL pathway loss renders ccRCCs highly glycolytic even under normoxia, leading to continuous production of lactate and protons. Because metabolite production scales with tumour volume, whereas clearance occurs across the tumour surface, larger tumours accumulate metabolic waste more rapidly. This geometric imbalance reduces the surface-area-to-volume (SA:Vol) ratio ($3/R$), limiting oxygen delivery and immune access while impairing clearance of lactate.

We therefore hypothesised that as tumours grow, they cross an immunometabolic tipping point: a transition from an immune-permissive state to an immune-exhausted state driven by altered diffusion geometry. The goal of this model is not to capture all microenvironmental complexity, but to test whether geometry alone can recapitulate the observed size-dependent shift in immune dysfunction and metastatic propensity.

**Model set-up**

For simplicity, the tumour is assumed to be a perfect sphere of radius $R$. We solve the model in 1D radial space. For each tumour radius $R$, the volume $Vol = \frac{4}{3}\pi R^3$ and surface area $SA = 4\pi R^2$ were used to calculate the surface area to volume (SA:Vol) ratio ($3/R$), which modulates boundary condition scaling (e.g., oxygen and cytotoxic immune availability, lactate clearance).

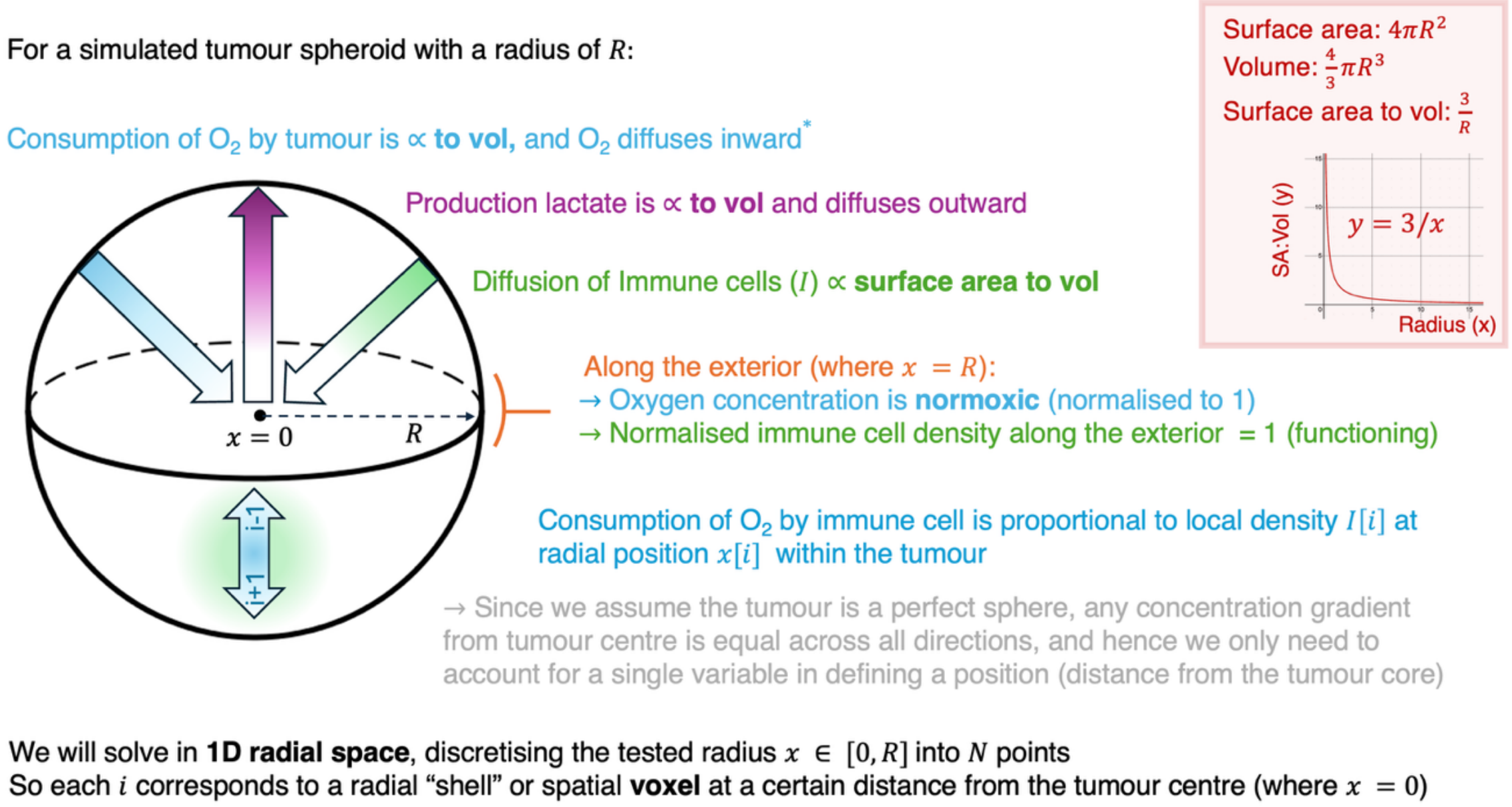

**Suppl. Fig. 3. Conceptual overview of model used.** Cytotoxic immune infiltration was assessed along a tumour spheroid for different tested radii, ranging from 0.5 cm to 5 cm. Since all tumour cells were assumed to be metabolically active, the tumour's total oxygen consumption was made proportional to the volume, while oxygen supply was assumed to diffuse inward, via the tumour surface. We assumed the tumour boundary is well-oxygenated/normoxic, as a proxy for exchange with surrounding tissue vasculature*. Production of lactate was assumed to be proportional to tumour volume as well, while diffusion of functional/cytotoxic immune cells was assumed to be proportional to the surface area to volume ratio (SA:Vol). Immune cells were assumed to consume oxygen locally.
*Unlike intratumoural vessels, the normal tissue surrounding the tumour retains functional perfusion, and thus provides a reliable source of oxygen diffusion at the tumour boundary. Tumours often coopt neighbouring vessels, further justifying oxygen diffusion from surrounding tissue into the tumour.

The radial domain was defined as $x \in [0, R]$, where $R$ is the tested tumour radius (tested from 0.5 – 5 cm). The domain was discretised into $N = 400$ evenly spaced points ($\Delta x$), and all spatial derivatives were approximated using central finite differences. For a radial coordinate $r = x[i]$, with step size $dx$, the following steady state equations were solved:

The following steady-state equations were solved iteratively for each metabolic 'species':

1) Lactate ($L(r)$, uniformly produced throughout tumour, diffusing outward);

$$\frac{1}{r^2}\frac{d}{dr}\left(r^2\frac{dL}{dr}\right) + \frac{p_L}{D_L} = 0 \qquad (S1)$$

Where $p_L$ is the lactate production rate, which is scaled with tumour volume, and $D_L$ is the lactate diffusion constant.

2) Immune cells ($I(r)$, diffuse inwards from tumour boundary, suppressed by lactate and/or hypoxia);

$$\frac{1}{r^2}\frac{d}{dr}\left(r^2\frac{dI}{dr}\right) - \frac{1}{D_I}\left(a_L L + a_O H(O)\right)I = 0 \qquad (S2)$$

Where $D_I$ is the immune cell diffusivity (inferred to be low due to size and e.g., stromal barriers), $a_L$ is the suppression due to lactate (/acidosis of co-secreted $H^+$), $a_O$ is the suppression due to true hypoxia, and $H(O)$ is the hypoxia indicator function, which is 1 if $O < O_{thresh}$ (hypoxia threshold), and 0 otherwise. $O_{thresh}$ was drawn from a uniform distribution $U(0.1,0.5)$, representing 10-50% of normoxic oxygen concentrations. Immune density $I(x)$ is a dimensionless variable representing the relative fraction of functional/cytotoxic immune density, where 1 represents healthy/active state, and 0 represents complete suppression/inactivation.

3) Oxygen ($O(r)$, diffuses from boundary inward, consumed by tumour and immune cells)

$$\frac{1}{r^2}\frac{d}{dr}\left(r^2\frac{dO}{dr}\right) - \frac{1}{D_O}\left(p_T + p_I I\right) = 0 \qquad (S3)$$

Where $D_O$ represents oxygen's diffusion (which is assumed to be faster than diffusion of the others), $p_T$ is oxygen consumption by tumour cells, which is scaled with tumour volume, while $p_I$ is local consumption of oxygen by immune cells (not scaled with volume but dependent on local $I$)

These PDEs were discretised using central finite differences on a radial mesh:

$$L_i = \frac{\left(r_{i+\frac{1}{2}}\right)^2 L_{i+1} + \left(r_{i-\frac{1}{2}}\right)^2 L_{i-1} + \frac{\Delta x^2 p_L r_i^2}{D_L}}{\left(r_{i+\frac{1}{2}}\right)^2 + \left(r_{i-\frac{1}{2}}\right)^2} \qquad (S4)$$

$$I_i = \frac{\left(r_{i+\frac{1}{2}}\right)^2 I_{i+1} + \left(r_{i-\frac{1}{2}}\right)^2 I_{i-1}}{\left(r_{i+\frac{1}{2}}\right)^2 + \left(r_{i-\frac{1}{2}}\right)^2 + \frac{\Delta x^2 r_i^2}{D_I}\left(a_L L_i + a_O H_i\right)} \qquad (S5)$$

$$O_i = \frac{\left(r_{i+\frac{1}{2}}\right)^2 O_{i+1} + \left(r_{i-\frac{1}{2}}\right)^2 O_{i-1} - \frac{\Delta x^2 r_i^2 (p_T + p_I I_i)}{D_O}}{\left(r_{i+\frac{1}{2}}\right)^2 + \left(r_{i-\frac{1}{2}}\right)^2} \qquad (S6)$$

Parameters were either fixed or randomly sampled:

1) Diffusion coefficients were set apart by orders of magnitude to represent relatively faster diffusion for smaller 'species', such that $D_O = 10, D_L = 1, D_I = 0.01$;
2) Immune cell suppression parameters were drawn from uniform distributions: $a_L | a_O \sim U(0.1, 0.9)$
3) Production rate of lactate and consumption rate of oxygen by the tumour scaled with volume, such that $p_L = p_{L0} Vol,\ p_T = p_{T0} Vol$

**Boundary & initial conditions**

Boundary conditions were informed by geometric constraints, such that:

1) Oxygen and cytotoxic immune cell access at the tumour boundary were scaled by the SA:Vol ratio, under the rationale that as tumours grow, a relatively smaller surface is 'responsible' for supplying oxygen and immune access to a growing mass;
$$O_{ext} = O_{base} \frac{SA}{Vol};\quad I_{ext} = I_{base} \frac{SA}{Vol} \qquad (S7)$$
We set both $O_{base}$ and $I_{base}$ to 1, representing normoxia and healthy/functional immune density, respectively.
2) Lactate clearance at the tumour boundary (i.e., peritumoural region) was scaled inversely with the SA:Vol ratio, to simulate impaired metabolite clearance in larger tumours;
$$L_{ext} = \max\left(0, 1 - \frac{SA}{Vol}\right) \qquad (S8)$$

This resulted in greater lactate accumulation in larger tumours, mimicking the biophysical constraint in diffusion and thus impaired lactate clearance in larger tumours.

Boundary conditions were implemented using ghost points with mirror reflection for the core (i.e., $L[1] = L[2]$, etc.).

For each tumour radius tested, the system was initialised with $L = 0, I = 0.1, O = 1$. Thus, the model's starting condition reflects no pre-existing lactate buildup by the tumour, low-level basal immune infiltration throughout the tumour upon initiation, and complete oxygenation throughout the tumour, respectively. Biologically, this models cytotoxic immune cells as already present or attempting to access the tumour, even if diffusion and suppression have not yet taken full effect. Iterative updates were then applied to solve the system until convergence.

**Numerical solving**

The steady-state equations were solved using a fixed-point iterative scheme on the discretised radial grid, updating each species until convergence (tolerance $<10^{-4}$). This approach is equivalent to integrating the diffusion-reaction equations forward in time to steady state but avoids explicit time stepping. All simulations were implemented in R (v 4.2.3).
Each tumour radius was simulated over 50 independent replicates with resampled parameters to account for parameter stochasticity (of those drawn from uniform distributions).
For comparison, we simulated control conditions where we assumed no hypoxia-induced or lactate-induced limitation on immune cell activity/migration/recruitment via $a_O = 0$ or $a_L = 0$, respectively.

**Outcome metric**

Then, for each round of integrations, we calculated a Immune Dysfunction Score (IDS), which was defined as the fraction of tumour volume where functional/cytotoxic immune cell density $I(x)$ fell below a critical threshold of $I_{thresh}$, which we set to 0.2, representing immune exhaustion/suppression (<20% functional cytotoxic cell density). This metric was computed across a range of tumour radii $R \in [0.5, 5.0]$ cm in 0.25 cm increments. The IDS fraction was summarised (median) across replicates.

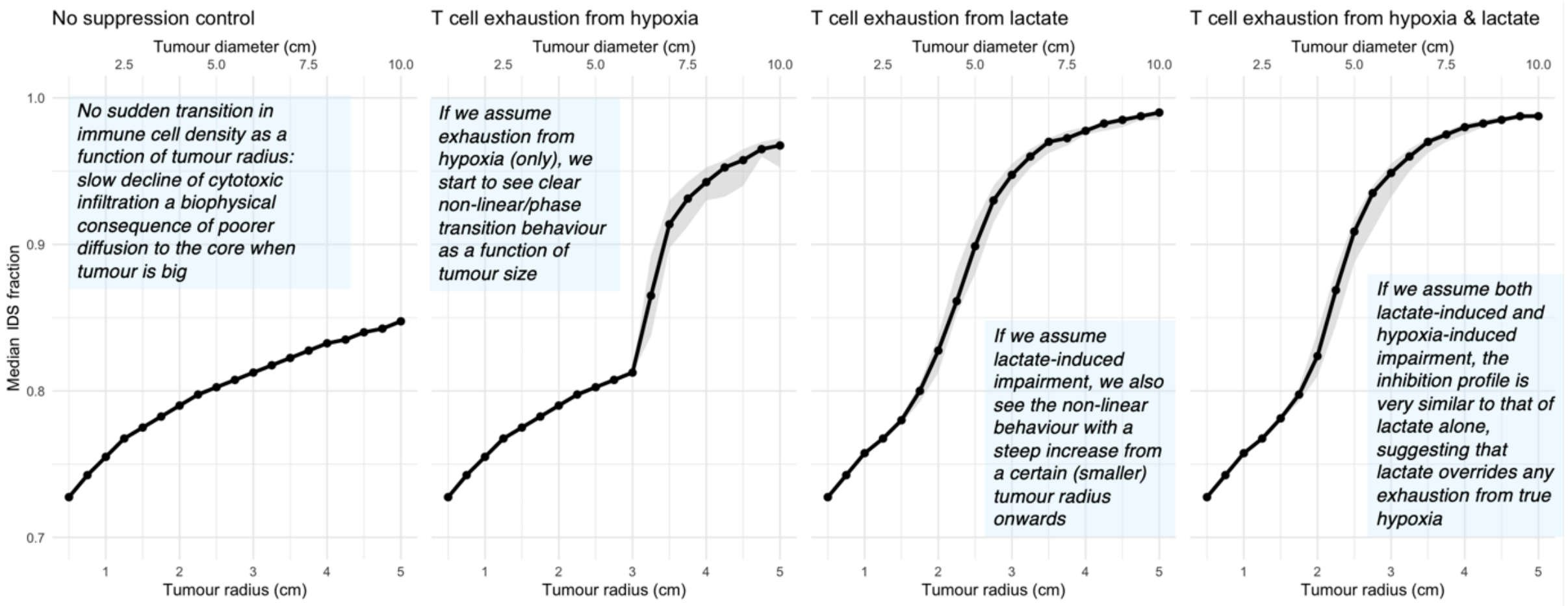

**Suppl. Fig. 4. Model outcomes for simulations with and without hypoxia-induced and/or lactate-induced immune cell exhaustion.**
Note that for the hypoxia penalty, due to the nature of the indicator function, the transition is more sudden compared to how the model accounts for lactate impairment (which was proportional to lactate concentration, whereas hypoxia-related T cell exhaustion occurs when oxygen drops below the hypoxia boundary).

**Results & Discussion**

The model predicts that as tumours enlarge, diffusion geometry alone can trigger a sharp transition from active immune function to immune exhaustion, driven by accumulation or impaired clearance of glycolytic byproducts such as lactate (**Suppl. Fig. 4**). This transition emerges without invoking new genetic alterations, suggesting that the clinical "3 cm rule" reflects a breakdown of immune-metabolic equilibrium rather than a mutational threshold. Within this framework, small tumours remain in a quasi-stable state where immune surveillance and metabolism are balanced, while larger tumours accumulate lactate and acidosis, generating an immunosuppressive niche that enables metastatic competence.

Importantly, the emergence of an exhaustion threshold occurred even when hypoxia was removed from the model, indicating that size-dependent lactate accumulation alone can generate an immune-suppressive niche. However, the model assumes spherical symmetry, uniform metabolic activity, steady-state conditions, and a single effective immune population (i.e., no explicit 'split' of immune cell population into immune exhausted population). It does not incorporate explicit angiogenesis, stromal barriers, or T cell-derived lactate, although the latter is qualitatively captured by the lactate-dependent immune suppression term.

Thus, results should be interpreted qualitatively, illustrating a plausible mechanistic basis by which tumour geometry alone can generate a size-dependent immunometabolic shift.